\documentclass[12pt,thmsa]{article}

\input tcilatex

\begin{document}

\title{\textbf{HIGH-TEMPERATURE ANYON SUPERCONDUCTIVITY}\\
} 
\author{\textit{E.J. Ferrer}\thanks{%
E-mail: ferrer@fredonia.edu},\textit{\ R. Hurka and V. de la Incera}$\thanks{%
E-mail: incera@fredonia.edu}$ \and Dept. of Physics, State University of New
York \and Fredonia, NY 14063, USA}
\date{SUNY-FRE-96-02}
\maketitle

\begin{abstract}
The screening of an applied magnetic field in a charged anyon fluid at
finite density ($\mu \neq 0$) and temperature ($T\neq 0$) is investigated.
For densities typical of high-temperature superconducting materials we find
that the anyon fluid exhibits a superconducting behavior at any temperature.
The total Meissner screening is characterized by two penetration lengths
corresponding to two short-range eigenmodes of propagation within the anyon
fluid.
\end{abstract}

Since the proposition by Laughlin and his collaborators\cite{1},\cite{2}
that condensation of bosonic composites of \textit{anyonic} quasiparticles
could give rise to high-T$_C$ superconductivity, a significant work has been
done in this direction.

The anyon superconductivity at $T=0$ has been investigated by many authors%
\cite{2}-\cite{11}. Crucial to the interpretation of anyon superconductivity
at $T=0$ was the exact cancellation between the bare and induced
Chern-Simons terms in the effective action of the theory.

In a recent paper\cite{13a}, it was stressed that a self-consistent
treatment of many-particle systems (i.e. $n_e\left( \mu \right) \neq 0$) at
zero temperature requires to consider the zero temperature statistical limit
($T\rightarrow 0,$ $\mu \neq 0$) instead of the QFT ($T=0,$ $\mu =0$)
formulation. In that case it was proven\cite{13a} that, contrary to what
occurs in quantum field theory ($T=0$), superconductivity arises only in the
case that an external electric field, transverse to the supercurrent in the
plane of the two-dimensional fluid, is considered. This transverse electric
field can be interpreted as simulating the possible effects of the
transverse voltage created by vortices\cite{4}.

The possible realization of anyon superconductivity at $T\neq 0$ has also
been extensively investigated\cite{8}-\cite{11},\cite{13},\cite{15}. At
finite temperature, based on non-vanishing corrections to the induced
Chern-Simons coefficient, some authors (see, ref. \cite{9}) have concluded
that the superconductivity is lost at $T\neq 0$. In contrast with this
result, in refs. \cite{11},\cite{15} it was argued that the non-vanishing
corrections to the induced Chern-Simons coefficient is numerically
negligible at $T\sim 100$ $^{\circ }K$. On the other hand, the development
of a pole $\sim \left( \frac 1{\mathbf{k}^2}\right) $ at $T\neq 0$ in the
polarization operator component $\Pi _{00}$, characteristic of the Debye
screening in plasmas, was found\cite{11},\cite{15} as the main reason for
the lack of a total Meissner effect in the charged anyon fluid at finite
temperature. In these papers it was discussed how the appearance of this
pole leads to a partial Meissner effect with a penetration which appreciably
increases with temperature. In ref. \cite{8}, similar results were
independently obtained. There, it was claimed that the anyon model fails to
provide a good superconducting behavior at finite temperature. The reason is
that three independent components of the magnetic interaction were obtained
within the charged anyon fluid at $T\neq 0$. Two of finite-range and a third
one of long-range which vanishes only at $T=0$.

In the present paper, working in the self-consistent field approximation
\cite{8},\cite{11},\cite{13a},\cite{15}, we will show that the charged anyon
fluid exhibits a superconducting behavior at any temperature. This
superconductivity is characterized by a total Meissner screening. The
external magnetic field is damped within the anyon fluid by two
characteristic lengths corresponding to two short-range eigenmodes of
propagation. We emphazise that the asymptotic conditions ($x\rightarrow
\infty $) for the zero components of the Maxwell and Chern-Simons gauge
potentials, $A_0$ and $a_0$ respectively, have been crucial to obtain this
behavior. These field components play the role of Lagrange multipliers in
the Extended Hamiltonian of the Dirac formulation of this theory. As it is
known, the restrictions in the allowed asymptotic behavior of the Lagrange
multipliers have profound consequences for gauge field theories\cite{19}.

The approach we follow is to compute the finite temperature effective action
in the self-consistent field approximation starting from the Lagrangian
density

\begin{equation}
\mathcal{L}=-\frac 14F_{\mu \nu }^2-\frac N{4\pi }\varepsilon ^{\mu \nu \rho
}a_\mu \partial _\nu a_\rho +en_eA_0+i\psi ^{\dagger }D_0\psi -\frac
1{2m}\left| D_k\psi \right| ^2+\psi ^{\dagger }\mu \psi  \label{1}
\end{equation}
of a 2+1 dimension charged fluid of non-relativistic electrons, $\psi $,
coupled to two independent gauge fields, $A_\mu $ and $a_\mu $, which
represent the electromagnetic field and the Chern-Simons field respectively.
The covariant derivative is given by $D_\nu =\partial _\nu +i\left( a_\nu
+eA_\nu \right) ,\quad \nu =0,1,2.$ The charged character of the fluid is
implemented through the chemical potential $\mu $; $n_e$ is a background
neutralizing ``classical'' charge density. From the electric charge
neutrality condition, it is found\cite{11},\cite{13a} that the system ground
state has a non-zero expectation value of the Chern-Simons magnetic field $%
\left( \overline{b}=\frac{2\pi n_e}N\right) $.

Integrating out the electron field, we obtain the effective action
corresponding to the Lagrangian density (\ref{1}),

\begin{equation}
\Gamma _{eff}\,\left( A_\nu ,a_\nu \right) =-\frac 14F_{\mu \nu }^2-\frac
N{4\pi }\varepsilon ^{\mu \nu \rho }a_\mu \partial _\nu a_\rho
+en_eA_0-\beta ^{-1}\ln \det G^{-1}\left( A_\nu ,a_\nu \right)  \label{2}
\end{equation}
Here $G\left( A_\nu ,a_\nu \right) $ denotes the exact fermion Green's
function in the many-particle background (i.e. in the presence of $\overline{%
b}$ and $\mu $), and $\beta $ is the inverse absolute temperature.

To investigate the linear response of the medium to an applied external
magnetic field, it is enough to consider small fluctuations of the gauge
potentials around the many-particle ground state. That is, we can evaluate $%
\Gamma _{eff}$ up to second order in these small quantities,

\begin{equation}
\Gamma _{eff}\,\left( A_\nu ,a_\nu \right) =-\frac 14F_{\mu \nu }^2-\frac
N{4\pi }\varepsilon ^{\mu \nu \rho }a_\mu \partial _\nu a_\rho
+en_eA_0+\Gamma ^{\left( 2\right) }  \label{3}
\end{equation}

\begin{eqnarray}
\Gamma ^{\left( 2\right) } &=&\int dx\Pi _\nu \left( x\right) \left[ a_\nu
\left( x\right) +eA_\nu \left( x\right) \right] +  \nonumber  \label{4} \\
&&+\int dxdy\left[ a_\nu \left( x\right) +eA_\nu \left( x\right) \right] \Pi
_{\mu \nu }\left( x,y\right) \left[ a_\nu \left( y\right) +eA_\nu \left(
y\right) \right]  \label{4}
\end{eqnarray}
$\Gamma ^{\left( 2\right) }$ is the fermion contribution to the effective
action in the above approximation, $\Pi _\nu $ and $\Pi _{\mu \nu }$
represent the fermion tadpole and polarization operators respectively. An
essential point in the study of this effective theory is the calculation of
these operators by using the fermion thermal Green's function defined in the
presence of the background field $\overline{b}$. With this aim, we apply the
Matsubara finite-temperature technique in the Landau gauge for the
electromagnetic field $A_\mu $, as well as for the Chern-Simons field $a_\mu
$, taking, as in ref. \cite{13a}, the thermal Green's function in the small
background density approximation $\left( n_e<Nm^2\right) $. This
approximation is satisfied within the order of the typical densities
reported in high-temperature superconductivity\cite{11}.

The leading behavior of these operators for static $\left( k_0=0\right) $
and slowly $\left( \mathbf{k}\sim 0\right) $ varying configurations, and
with the spatial momentum specialized in the frame $\mathbf{k=}\left(
k,0\right) $, are (the details of the calculations will be published
elsewhere)

\begin{equation}
\Pi _k\left( x\right) =0,\qquad \qquad \Pi _4\left( x\right) =-n_e  \label{5}
\end{equation}

\begin{equation}
\Pi _{\mu \nu }=\left(
\begin{array}{ccc}
\mathit{\Pi }_{\mathit{0}}+\mathit{\Pi }_{\mathit{0}}\,^{\prime }\,k^2 & 0 &
\mathit{\Pi }_{\mathit{1}}k \\
0 & 0 & 0 \\
-\mathit{\Pi }_{\mathit{1}}k & 0 & \mathit{\Pi }_{\,\mathit{2}}k^2
\end{array}
\right)  \label{6}
\end{equation}

where

\begin{equation}
\mathit{\Pi }_{\mathit{0}}=-\frac m{4\pi }\left[ \tanh \left( \frac{\beta
\mu }2\right) +1\right]  \label{7}
\end{equation}

\begin{equation}
\mathit{\Pi }_{\mathit{0}}\,^{\prime }=-\frac \beta {48\pi }\limfunc{sech}{}%
^2\left( \frac{\beta \mu }2\right)   \label{8}
\end{equation}

\begin{equation}
\mathit{\Pi }_{\mathit{1}}=\frac{i\beta \overline{b}}{96\pi m}\limfunc{sech}{%
}^2\left( \frac{\beta \mu }2\right)   \label{9}
\end{equation}

\begin{equation}
\mathit{\Pi }_{\,\mathit{2}}=\frac 1{48\pi m}\left[ \tanh \left( \frac{\beta
\mu }2\right) +1\right]  \label{10}
\end{equation}
The operators (\ref{5}), and (\ref{6}) reduce to the ones reported in ref.
\cite{11} in the $T\rightarrow 0$ limit.

To investigate the linear response of the anyon fluid to an applied external
magnetic field we have to find the extremum equations derived from the
effective action (\ref{3}). This formulation is what is known in the
literature as the self-consistent field approximation\cite{11},\cite{15}. As
our main interest is to investigate the behavior, within the anyon fluid, of
an applied external magnetic field, we have to define a geometry for the
medium. As usual, we will consider a material confined to a semi-infinite
plane $-\infty <y<\infty $ with boundary at $x=0$. The external magnetic
field will be applied from the vacuum ($-\infty <x<0$). We restrict our
solution to gauge field configurations which are static and uniform in the $%
y $-direction.

The corresponding Maxwell and Chern-Simons extremum equations are
respectively,

\begin{equation}
\partial _\nu F^{\nu \mu }=eJ_{ind}^\mu  \label{11}
\end{equation}

\begin{equation}
-\frac N{4\pi }\varepsilon ^{\mu \nu \rho }f_{\nu \rho }=J_{ind}^\mu
\label{12}
\end{equation}
Here, $f_{\mu \nu }$ is the Chern-Simons gauge field strength tensor,
defined as $f_{\mu \nu }=\partial _\mu a_\nu -\partial _\nu a_\mu $, and $%
J_{ind}^\mu $ is the 4-current density induced by the anyon system at finite
temperature and density. Their different components are given by

\begin{equation}
J_{ind}^4\left( x\right) =\mathit{\Pi }_{\mathit{0}}\left[ a_0\left(
x\right) +eA_0\left( x\right) \right] +\mathit{\Pi }_{\mathit{0}}\,^{\prime
}\partial _x\left( \mathcal{E}+eE\right) +i\mathit{\Pi }_{\mathit{1}}\left(
b+eB\right)  \label{13}
\end{equation}

\begin{equation}
J_{ind}^1\left( x\right) =0,\qquad J_{ind}^2\left( x\right) =i\mathit{\Pi }_{%
\mathit{1}}\left( \mathcal{E}+eE\right) +\mathit{\Pi }_{\,\mathit{2}%
}\partial _x\left( b+eB\right)  \label{14}
\end{equation}
in the above expressions we used the following notation: $\mathcal{E}=f_{01}$%
, $E=F_{01}$, $b=f_{21}$ and $B=F_{21}$. Eqs. (\ref{13})-(\ref{14}) play the
role in the anyon fluid of the London equations in BCS superconductivity.
When the induced currents (\ref{13}), (\ref{14}) are substituted in eqs. (%
\ref{11}) and (\ref{12}) we find, after some manipulation, the set of
independent differential equations,

\begin{equation}
\omega \partial _x^2B+\alpha B=\gamma \left[ \partial _xE-\sigma A_0\right]
+\tau \,a_0  \label{15}
\end{equation}

\begin{equation}
\partial _xB=\kappa \partial _x^2E+\eta E  \label{16}
\end{equation}

\begin{equation}
\partial _xa_0=\chi \partial _xB  \label{17}
\end{equation}
The coefficients appearing in these differential equations depend on the
components of the polarization operators through the relations,

\[
\omega =\frac{2\pi }N\mathit{\Pi }_{\mathit{0}}\,^{\prime },\quad \alpha
=ie^2\mathit{\Pi }_{\mathit{1}},\quad \tau =-e\mathit{\Pi }_{\mathit{0}%
},\quad \chi =-\frac{2\pi }{eN},\quad \sigma =\frac{e^2}\gamma \mathit{\Pi }%
_{\mathit{0}},\quad \eta =-\frac{ie^2}\delta \mathit{\Pi }_{\mathit{1}}
\]

\begin{equation}
\gamma =1-e^2\mathit{\Pi }_{\mathit{0}}\,^{\prime }-\frac{2\pi i}N\mathit{%
\Pi }_{\mathit{1}},\quad \delta =1+e^2\mathit{\Pi }_{\,\mathit{2}}+\frac{%
2\pi i}N\mathit{\Pi }_{\mathit{1}},\quad \kappa =-\frac{2\pi }{N\delta }%
\mathit{\Pi }_{\,\mathit{2}},  \label{18}
\end{equation}

The extremum equations (\ref{15})-(\ref{17}) are not essentially different
from those found for the anyon effective theory at finite temperature by
other authors\cite{11},\cite{15}. Distinctive of these equations is the
appearance of the nonzero constant coefficients $\sigma $ and $\tau $. They
are related to the Debye screening which is a property of any charged
medium. It is a peculiar fact that in the anyon fluid these coefficients
appear linked to the magnetic field (see eq. (\ref{15})). As a consequence,
the zero components of the gauge potentials, $A_0$ and $a_0$, play a
nontrivial role in the field equations for the magnetic field. Therefore, it
is natural to expect that their contribution to the Meissner effect can be
significant. As we will show below, to consider the proper asymptotic
behavior of these potentials is crucial for the realization of the total
Meissner screening in the charged anyon fluid.

To solve eqs. (\ref{15})-(\ref{17}) we can conveniently arrange them to
obtain,

\begin{equation}
a\partial _x^4E+d\partial _x^2E+cE=0  \label{19}
\end{equation}
where $a=\omega \kappa $, $d=\omega \eta +\alpha \kappa -\gamma -\tau \kappa
\chi $, and $c=\alpha \eta -\sigma \gamma -\tau \eta \chi $. Then the
solutions for the fields $E$, and $B$, and for the potentials $a_0$ and $A_0$%
, can be obtained from (\ref{19}), (\ref{16}), (\ref{17}) and the definition
of $E$ in terms of $A_0$, respectively. Being (\ref{19}) a higher order
differential equation, its solution belongs to a wider class if compared to
that corresponding to the original eqs. (\ref{15})-(\ref{17}). Thus, to
exclude the redundant solutions we have to require that they satisfy eq. (%
\ref{15}) as a supplementary condition. In this way we can reduce the number
of independent unknown coefficients to six, which is the number
corresponding to the original system (\ref{15})-(\ref{17}).

Solving eq. (\ref{19}) we obtain,

\begin{equation}
E\left( x\right) =C_1e^{-x\xi _1}+C_2e^{x\xi _1}+C_3e^{-x\xi _2}+C_4e^{x\xi
_2},  \label{20}
\end{equation}
where

\begin{equation}
\xi _{1,2}=\left[ -d\pm \sqrt{d^2-4ac}\right] ^{\frac 12}/\sqrt{2a}
\label{20a}
\end{equation}
take real values at any temperature when evaluated with the typical values $%
n_e=\left( 1\sim 5\right) \times 10^{14}cm^{-2}$, $m=2m_e$ ($m_e=2.6\times
10^{10}cm^{-1}$ is the electron mass) and $\left| N\right| =2$.

With the solution (\ref{20}), eqs. (\ref{16}), (\ref{17}), and the
definition $E=-\partial _xA_0$, we find,

\begin{equation}
B\left( x\right) =\frac{\xi _1^2\kappa +\eta }{\xi _1}\left( C_1e^{-x\xi
_1}-C_2e^{x\xi _1}\right) +\frac{\xi _2^2\kappa +\eta }{\xi _2}\left(
C_3e^{-x\xi _2}-C_4e^{x\xi _2}\right) +C_5  \label{21}
\end{equation}

\begin{equation}
a_0\left( x\right) =\chi \frac{\xi _1^2\kappa +\eta }{\xi _1}\left(
C_1e^{-x\xi _1}-C_2e^{x\xi _1}\right) +\chi \frac{\xi _2^2\kappa +\eta }{\xi
_2}\left( C_3e^{-x\xi _2}-C_4e^{x\xi _2}\right) +C_6  \label{22}
\end{equation}

\begin{equation}
A_0\left( x\right) =\frac 1{\xi _1}\left( -C_1e^{-x\xi _1}+C_2e^{x\xi
_1}\right) +\frac 1{\xi _2}\left( -C_3e^{-x\xi _2}+C_4e^{x\xi _2}\right) +C_7
\label{23}
\end{equation}
The extra unknown coefficient is eliminated, as it was explained above,
substituting the solutions (\ref{20}), (\ref{21}), (\ref{22}) and (\ref{23})
into eq. (\ref{15}) to obtain the relation,

\begin{equation}
C_5=\frac \tau \alpha C_6+\frac{\sigma \gamma }\alpha C_7  \label{24}
\end{equation}
The last relation establishes a connection between the asymptotic conditions
for the zero components of the gauge potentials and the asymptotic condition
for the magnetic field.

To determine the six independent coefficients we have to consider the proper
boundary conditions for the possible realization of the Meissner effect.
Thus, we consider that the magnetic field has a boundary value $B\left(
x=0\right) =\overline{B}$ and is finite when $x\rightarrow \infty $, that
the boundary value of the electric field is zero, $E\left( x=0\right) =0$,
and that the asymptotic values of the zero component of the gauge potentials
are zero, i.e. $A_0\left( x\rightarrow \infty \right) =0$ and $a_0\left(
x\rightarrow \infty \right) =0$. These asymptotic conditions for the gauge
potentials guarantee well behaved Poisson brackets between the dynamical
variables of the theory and the generators of the gauge transformations\cite
{19}.

With the above conditions we obtain that $C_2=C_4=C_5=C_6=C_7=0$ and $%
C_1=-C_3$, where $C_1$ depends on the magnetic field boundary value and
temperature through the relation,

\begin{equation}
C_1=\frac{\xi _1\xi _2}{\left( \xi _2^2\kappa +\eta \right) -\left( \xi
_1^2\kappa +\eta \right) }\overline{B}  \label{25}
\end{equation}
The magnetic field penetration (i.e. at $x\geq 0$) is then given by,

\begin{equation}
B\left( x\right) =B_1\left( T\right) \,e^{-x\xi _1}+B_2\left( T\right)
\,e^{-x\xi _2}  \label{26}
\end{equation}
where the temperature dependent coefficients are,

\begin{equation}
B_1\left( T\right) =\frac{\xi _2\left( \xi _1^2\kappa +\eta \right) }{\left(
\xi _1^2\kappa +\eta \right) -\left( \xi _2^2\kappa +\eta \right) }\overline{%
B},\qquad B_2\left( T\right) =\frac{\xi _1\left( \xi _2^2\kappa +\eta
\right) }{\left( \xi _2^2\kappa +\eta \right) -\left( \xi _1^2\kappa +\eta
\right) }\overline{B}  \label{27}
\end{equation}

Hence, we have that within the anyon fluid the applied magnetic field falls
down exponentially on two essentially different scales, $\lambda _1=1/\xi _1$
and $\lambda _2=1/\xi _2$, which characterize two eigenmodes of propagation
inside the fluid. If we define the effective penetration length, $\overline{%
\lambda }$, as the distance $x$ where the magnetic field falls down to a
value $B\left( \overline{\lambda }\right) /\overline{B}\simeq 0.38$, we find
that at $T\approx 200K$ the screening is complete at $\overline{\lambda }%
\approx 8.5\times 10^{-11}$ $cm$. Considering the obtained values for the $%
C_i$'s coefficients in the solution (\ref{20}), we also find that the
induction of an electric field is intimately linked to the Meissner effect
in the anyon fluid.

Finally, a few comments are in order. We want to point out that a partial
Meissner solution\cite{11},\cite{15}, would imply a variation in the
asymptotic conditions of the gauge potentials. Specifically, it can be
obtained for $A_0\left( x\rightarrow \infty \right) =\overline{D}_1$ and $%
a_0\left( x\rightarrow \infty \right) =\overline{D}_2$, ($\overline{D}_1$
and $\overline{D}_2$ constants). In statistical gauge theory these constant
asymptotic field configurations are not gauge equivalent (under proper,
periodic gauge transformations) to the trivial vacuum\cite{20}. To study the
problem under these nontrivial asymptotic conditions we have to reconsider
the whole gauge formulation of the theory taking into account the possible
appearance of an overall ''charge rotation''\cite{19}. An important point
that we will discuss in detail in a future publication is the existence of
some indications, within the present approximation, of possible phase
transitions from the superconducting phase to the normal one.

This research has been supported in part by the National Science Foundation
under Grant No. PHY-9414509.

\end{document}